\documentclass{emulateapj}
\usepackage{graphicx}
\usepackage{amsmath}

\slugcomment{ApJL accepted}

\shorttitle{Surface Retrieval on Exoplanets}
\shortauthors{Cowan \& Strait}

\begin{document}

\title{Determining Reflectance Spectra of Surfaces and Clouds on Exoplanets}

\author{Nicolas B. Cowan\altaffilmark{1,2}, Talia E. Strait\altaffilmark{2,1}}
\altaffiltext{1}{Center for Interdisciplinary Exploration and Research in Astrophysics (CIERA), Northwestern University, 2131 Tech Dr., IL 60208, USA}
\altaffiltext{2}{Department of Physics and Astronomy, Northwestern University, 2145 Sheridan Rd., F165, Evanston, IL 60208, USA}

\email{n-cowan@northwestern.edu}

\begin{abstract}
Planned missions will spatially resolve temperate terrestrial planets from their host star.  Although reflected light from such a planet encodes information about its surface, it has not been shown how to establish surface characteristics of a planet without assuming known surfaces to begin with. We present a re-analysis of disk-integrated, time-resolved, multiband photometry of Earth obtained by the Deep Impact spacecraft as part of the EPOXI Mission of Opportunity.  We extract reflectance spectra of clouds, ocean and land without a priori knowledge of the numbers or colors of these surfaces.  We show that the inverse problem of extracting surface spectra from such data is a novel and extreme instance of spectral unmixing, a well-studied problem in remote sensing. Principal component analysis is used to determine an appropriate number of model surfaces with which to interpret the data. Shrink-wrapping a simplex to the color excursions of the planet yields a conservative estimate of the planet's endmember spectra. The resulting surface maps are unphysical, however, requiring negative or larger-than-unity surface coverage at certain locations.  Our ``rotational unmixing'' supersedes the endmember analysis by simultaneously solving for the surface spectra and their geographical distributions on the planet, under the assumption of diffuse reflection and known viewing geometry. We use a Markov Chain Monte Carlo to determine best-fit parameters and their uncertainties. The resulting albedo spectra are similar to clouds, ocean and land seen through a Rayleigh-scattering atmosphere. This study suggests that future direct-imaging efforts could identify and map unknown surfaces and clouds on exoplanets. 
\end{abstract}

\keywords{methods: data analysis --- techniques: photometric --- planets and satellites: surfaces --- planets and satellites: atmospheres --- planets and satellites: individual (Earth)}

\section{Introduction}
Next-generation space telescopes will spatially resolve terrestrial exoplanets from their host star \citep{Traub_2011}. The rotational color variations of a ``pale blue dot'' might betray the presence of landforms rotating in and out of view \citep{Ford_2001}. Ground truth is critical to our understanding of planetary climate because surface liquid water is the definition of habitability \citep{Abe_1993, Kasting_1993} and long-term habitability may require exposed continents \citep{Abbot_2012}. 

Previous research has shown that rotational color variations of a planet can yield rotation rate \citep{Palle_2008, Oakley_2009}, coarse planetary maps under the assumption of known surfaces \citep{Fujii_2010, Fujii_2011, Kawahara_2010}, single-band maps \citep{Kawahara_2011, Fujii_2012},  and maps of eigencolors \citep{Cowan_2009, Cowan_2011}.  We cannot assume, however, that terrestrial exoplanets will have the same surface types as Earth, and there is no obvious relation between single-band albedo or eigencolors with surface features, so neither of these ``exo-cartographic'' strategies is entirely satisfactory.  

Three methods have been proposed for using orbital phase variations to detect surface liquid water on exoplanets, but they have challenges of their own: thermal inertia \citep[thermal infrared;][]{Gaidos_2004, Cowan_2012c}, glint \citep[visible--near infrared;][]{Williams_2008, Robinson_2010, Cowan_2012b} and polarization \citep[visible--near infrared;][]{Zugger_2010, Zugger_2011}. Significantly, these methods leverage variability on the planet's orbital, rather than rotational, timescale.  While this makes the signal-to-noise requirements less stringent than exo-cartography, none of these methods would be able to distinguish between partial vs.\ complete ocean coverage. 

In this Letter, we show how rotational color variations can be used to retrieve reflectance spectra of a planet's dominant surfaces. Our method does not require a priori knowledge of the number of surfaces, let alone their colors or geographical distributions.

\section{Analysis}
We re-analyze disk-integrated Earth observations obtained by the Deep Impact spacecraft as part of the EPOXI Mission of Opportunity \citep{Livengood_2011}. Our data consist of hourly measurements of apparent albedo, $A^*$, in seven wavebands spanning a single rotation of Earth (the \emph{Earth5} time-series). The first and last observations occur exactly 24 hours apart but have slightly different apparent albedos because of day-to-day changes in cloud cover \citep{Goode_2001, Palle_2004, Cowan_2009, Cowan_2011}.  Specifically, given the orbital phase near quadrature, the final 6 observations probe the same regions as the early part of the time-series. 

\begin{figure}[htb]
      \includegraphics[width=80mm]{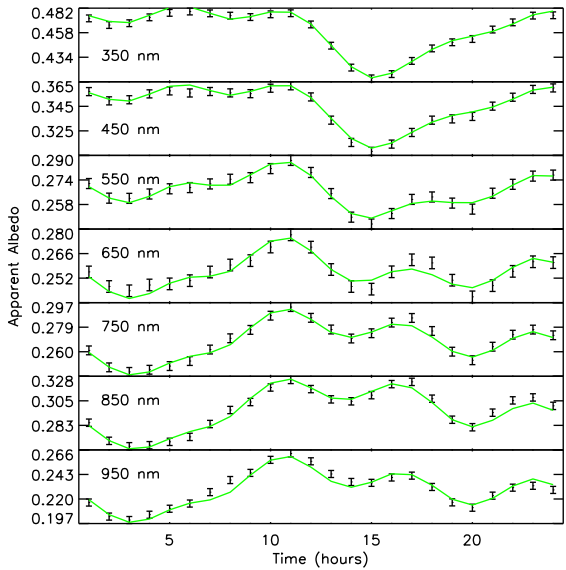}
    \caption{{\small Changes in the apparent albedo of Earth in seven wavebands spanning near-UV to near-IR over 24 hrs, as observed by the Deep Impact spacecraft as part of the EPOXI mission on Jun 4, 2008. The black points show the data with our adopted uncertainties. The green lines show our best-fit model with 9 longitudinal slices and 3 surface types. \label{lightcurves}}}
\end{figure}

We ``correct'' the final 6 observations so that the lightcurves may be fit with a static model. At each wavelength, we subtract the last datum from the first to estimate the effect of changing cloud cover. We then correct the final 6 observations according to this change: the $19^{\rm th}$ observation is unaffected, and the correction increases linearly from the $20^{\rm th}$ to the $25^{\rm th}$ observation. After the application of this correction, the last observation is identical to the first, by construction.  In order to avoid giving this point too much weight, we remove the final datum, resulting in 24 observations populating a 7-dimensional color space (Figure~\ref{lightcurves}).

\subsection{Principal Component Analysis}
We center the data by subtracting the planet's time-averaged color and then perform principal component analysis (PCA) using the covariance of the data \citep{Cowan_2009, Cowan_2011}.  The first two principal components have nearly equal power and account for 99\% of the color variance in our data (90\% of the variability).

\begin{figure*}[htb]
\begin{center}
$\begin{array}{cc}
      \includegraphics[width=80mm]{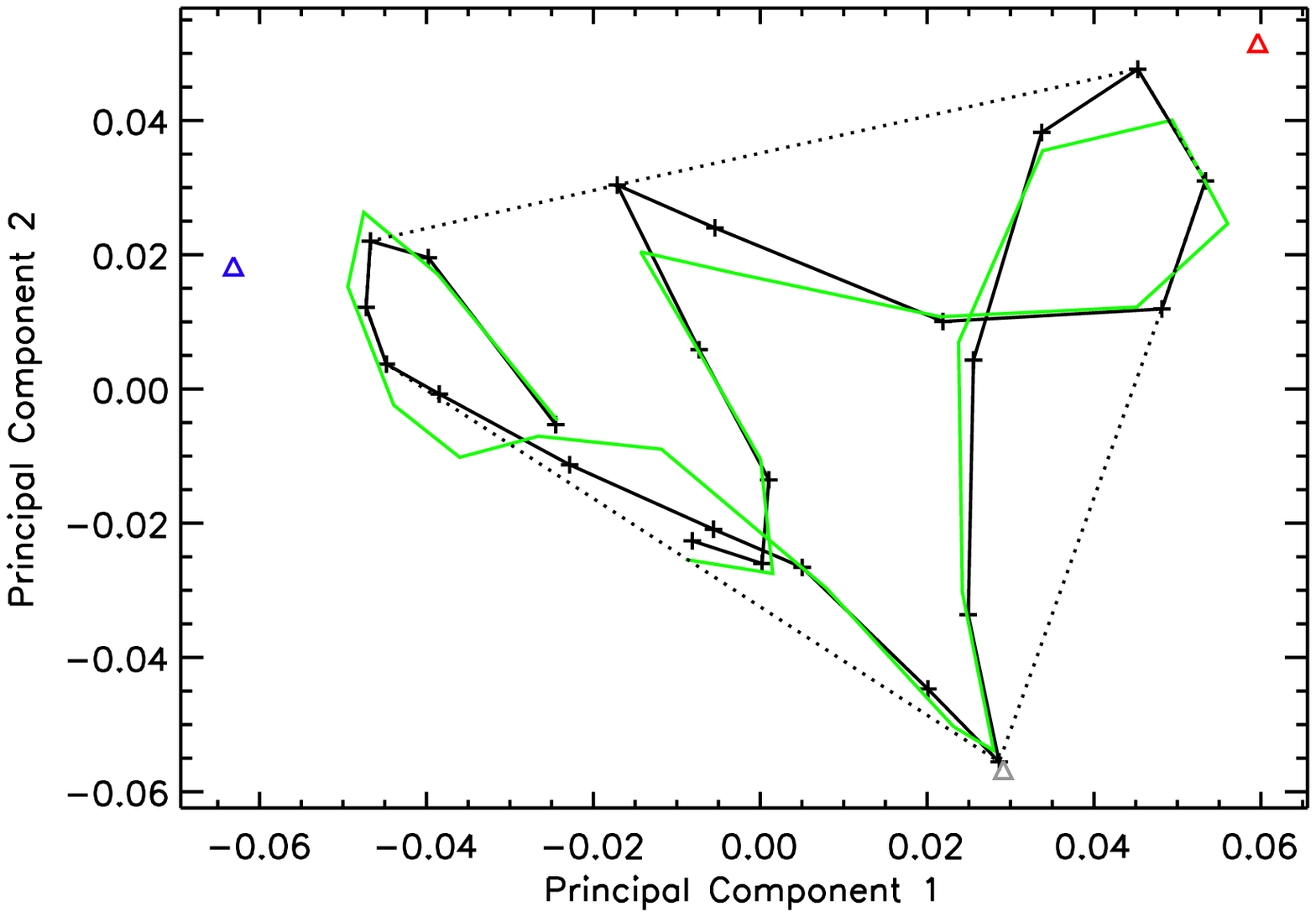}&
      \includegraphics[width=80mm]{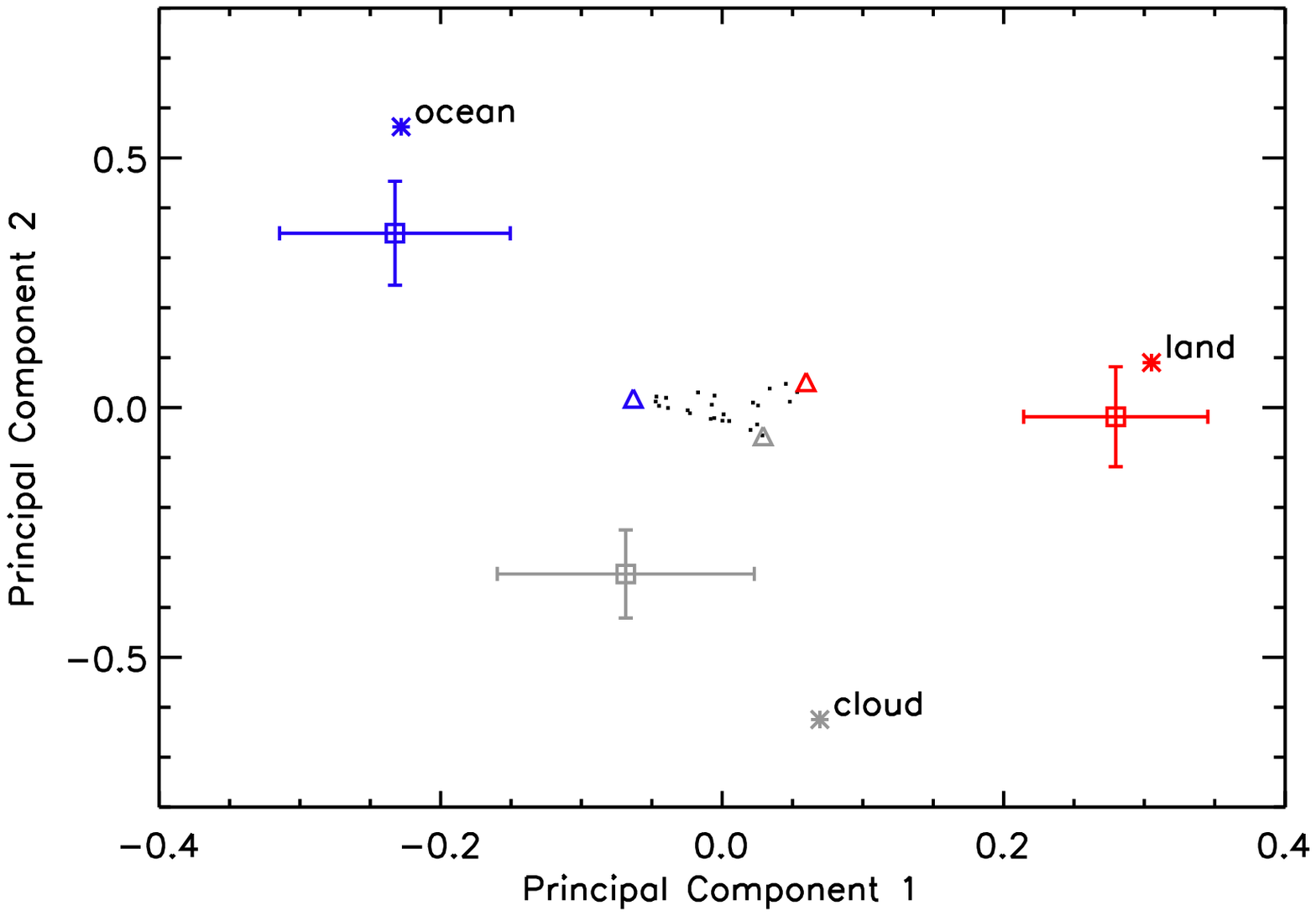}
\end{array}$ 
\end{center}
\caption{{\small Principal Component Plane. \emph{Left:} The black crosses connected by the solid black line show the EPOXI data projected onto the principal component plane. The dotted black line shows the convex hull of the observations. The colored triangles show the endmember spectra obtained by shrink-wrapping a simplex (a triangle, in this case) onto the convex hull. The green line shows the color excursions of our best-fit model obtained by rotational unmixing. \emph{Right:} The black points are the EPOXI data after zooming out. The colored triangles are the same endmember spectra as on the left. The colored squares with error bars are the surface spectra retrieved from the data by rotational unmixing. The colored asterisks show predicted spectra of ocean, land and cloud combined with Rayleigh scattering \citep{Robinson_2011, Cowan_2011}. \label{pcp}}}
\end{figure*}

There is no one-to-one correspondence between surface types and principal components, however: the latter are normalized and orthogonal and therefore unphysical \citep{Cowan_2011}. Furthermore, the number of dominant principal components does not even correspond to the minimum number of surfaces for reflected light surface retrieval, rather, $n_{\rm surf} \gtrsim n_{\rm dom}+1$ \citep{Cowan_2011}.  Nevertheless, it is useful to reduce the dimensionality of the data by projecting them onto the principal component plane (PCP). In the present case, the PCP is defined by the first two principal components (Figure~\ref{pcp}); in general it is a hyper-plane.
   
\subsection{Simplex Shrink-Wrapping}   
Insofar as the color excursions of Earth lie in the PCP, then so must its dominant surfaces. Since three point uniquely define a plane, Occam's razor dictates that we try a three-surface solution. If one assumes that different surfaces and regions combine linearly to determine the planet's over-all colors, then the three pure surface spectra must define a triangle that encloses the data.  

Estimating broadband endmember spectra from our locus of data is isomorphic to unsupervised spectral unmixing of multi-spectral satellite images \citep{Sabol_1992, Keshava_2003}. In traditional remote sensing, the data consists of colors at each pixel in a spatially-resolved image; in the present exoplanet  application, the data consist of colors at each point in time, where there is a non-trivial convolution tying the time-series to spatial inhomogeneities on the surface of the planet.  

We adopt a remote sensing algorithm: simplex shrink-wrapping \citep{Fuhrmann_1999}.  Recall that a simplex in $N$-dimensional space is a polyhedron with $N+1$ vertices: on a line the simplex is a line segment, on a plane the simplex is a triangle, in three-dimensions the simplex is a tetrahedron, etc.  
   
Simplex shrink-wrapping entails finding the simplex with the smallest volume (length in 1D; area in 2D) that encloses all the data. Since only the most extreme colors constrain the shrink-wrapping, it is expedient to first determine the convex hull of the locus in the PCP (the dotted black line in left panel of Figure~\ref{pcp}). This step is not computationally necessary for our small ($24\times7$) data matrix, but may be necessary for longer observations and/or higher spectral resolution.  
   
The apparent albedo can be modeled as the matrix product of surface spectra, $S$, and apparent covering fraction, $f^*$:  $A^*[n_{\rm t}, n_\lambda] = f^*[n_{\rm t}, n_{\rm surf}] \times S[n_{\rm surf}, n_\lambda]$, where $n_{\rm t}$ is the number of observations and $n_\lambda$ is the number of wavebands. Insofar as the observations do not lie perfectly in the principal component plane, this model cannot perfectly match the data. Instead, the object is to match the projected albedo in the PCP. Simplex shrink-wrapping amounts to solving for $f^*$ and $S$ given $A^*$ and the constraint of minimum volume. The vertices of our shrink-wrapped triangle are denoted by triangles in Figure~\ref{pcp}. The shrink-wrap endmembers are more likely to be unique if the data locus is non-spherical. If one knew nothing about the planet's orbital plane or viewing geometry, endmembers would be the most conservative estimate of surface spectra.
      
The apparent covering fraction is related to the planetary covering fraction, $f$, by the convolution: $f^*[n_{\rm t}, n_{\rm surf}] = W[n_{\rm t}, n_{\rm slice}] \times f[n_{\rm slice}, n_{\rm surf}]$, where $n_{\rm slice}$ is the number of longitudinal slices on the planet. The weight, $W$, quantifies the visibility and illumination of a given longitudinal slice on the planet at a given point in time. It is computed given the known sub-observer and sub-stellar positions at the time of each observation \citep{Cowan_2011}. Although we adopt a spatial resolution of $n_{\rm slice}=9$, the $W$ and $f$ arrays are oversampled by a factor of 100 in order to reduce numerical integration error.

Since $W$ is a known low-pass filter, acceptable $f^*$ (between zero and unity, and summing to unity at each point in time) may require unphysical $f$.  Indeed, if we adopt the shrink-wrap endmember spectra as bona fide surface spectra, the deconvolution, $f^* \to f$, produces surface maps with covering fractions greater than unity, or negative.  Since there is no possible surface map of the endmembers that matches the data, the endmembers must not correspond to the colors of actual surfaces. In other words, the shrink-wrap produces albedo spectra that are too conservative, and geographies that are too liberal. 

\subsection{Rotational Unmixing}
The essential difference between the present application and traditional spectral unmixing is that while the surface coverage, and hence colors, of a pixel may be arbitrarily different from that of its neighbor, the colors of a disk-integrated planet cannot change arbitrarily fast.
 
The third and final step of our analysis, after PCA and shrink-wrapping, is to solve for surface spectra, $S$, and geographies, $f$, given $A^*$ and $W$:  $A^*[n_{\rm t}, n_\lambda] = W[n_{\rm t}, n_{\rm slice}] \times f[n_{\rm slice}, n_{\rm surf}] \times S[n_{\rm surf}, n_\lambda]$.  This amounts to fitting the path of the planet in color-space by simultaneously varying the three surface spectra and their planetary geography. While simplex shrink-wrapping only uses the convex hull of the data, rotational unmixing makes use of all the observations.

We use a 3,000,000-step Markov Chain Monte Carlo (MCMC) to determine the best parameters and their uncertainties \citep{Ford_2001}.  We initialize the surface spectra at the endmember spectra from the shrink-wrapping to speed up convergence, and begin with a uniform surface map with equal amounts of the three surfaces.   Before evaluating a step in the MCMC, we ensure that the surface spectra have albedos between zero and unity at all wavelengths\footnote{Note that in practice it may be difficult to measure the absolute albedo of directly imaged terrestrial planets.}, and that the covering fractions are between zero and unity at every location on the planet. 
   
The MCMC sequentially takes seven 1D steps along each principal color for each surface spectrum ($n_{\rm surf}\times n_\lambda = 21$ steps), and a 3D step in map space for each longitudinal slice of the planet, normalized such that the sum of covering fractions is unity at each slice ($n_{\rm slice} = 9$ steps). Step sizes are tuned to get an acceptance ratio of $\sim0.25$ \citep{Gelman_2003}. We adopt Gaussian uncertainties of 0.0033 on the apparent albedo measurements, which produce a best-fit reduced $\chi^2$ of unity \citep[essentially the same 1\% uncertainties as in][]{Cowan_2009}. 
   
The surface spectra are not constrained to lie in the PCP, but the best fit solutions are close to the plane, as expected from PCA. Quantitatively, the surface spectra have 5-dimensional Euclidean distances from the principal component plane of 0.008, 0.007, and 0.009, for an aspect ratio of $\sim 2\%$ (cf. right panel of Figure~\ref{pcp}).  

Following \cite{Cowan_2009} and as dictated by the orbital phase, we use 9 longitudinal slices on the planet ($40^{\circ}$ spatial resolution, with fixed longitudes for the slices), convolving the surface map with the instantaneous illumination and visibility functions assuming diffuse (Lambertian) reflection and known viewing geometry (rotation rate, obliquity, orbital and seasonal phases).  Since the covering fractions for the three surfaces must sum to unity at each location on the planet, the model has $(n_{\rm surf} -1)n_{\rm slice} + n_{\rm surf} n_\lambda = 39$ free parameters.
   
\begin{figure}[htb]
      \includegraphics[width=80mm]{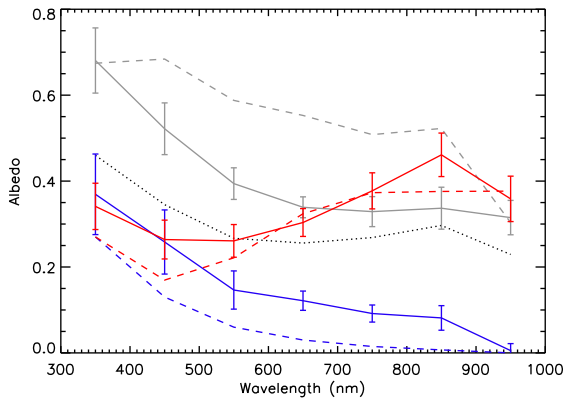}
    \caption{{\small Surface spectra of Earth as determined by the EPOXI observations. The solid colored lines show the best-fit surface spectra obtained by rotational unmixing; the colored dashed lines are the predicted spectra. The colors are the same as in Figure~\ref{pcp} and correspond to clouds (gray), ocean (blue), and land (red) viewed through a Rayleigh scattering atmosphere. The dotted black line shows the time-averaged broadband albedo spectrum of Earth. \label{spectra}}}
\end{figure}

The green lines in Figure~\ref{lightcurves} show the lightcurves for our best-fit model. The green line in the left panel of Figure~\ref{pcp} shows the centered color variations of the model projected on the PCP. The gray, red, and blue colored squares with error bars in Figure~\ref{pcp} denote our best-fit surface spectra projected onto the PCP. The three spectra correspond roughly to clouds, land, and ocean; their deprojected albedo spectra are shown in Figure~\ref{spectra}.  
   
In order to gauge the accuracy of the retrieval, we take empirical reflectance spectra of clouds, ocean and land from \cite{Robinson_2011} and combine them with an empirical model of disk-integrated Rayleigh scattering \citep{Cowan_2011}. The predicted surface spectra are the colored asterisks in Figure~\ref{pcp}, and the colored dashed lines in Figure~\ref{spectra}.
  
The qualitative agreement between the retrieved and predicted surface spectra is remarkable when one considers that the surface spectra are moving targets. Ocean, land and clouds on Earth are hardly uniform, and the path-length of Rayleigh scattering is a function of location on the disk of the planet. We have incorporated Rayleigh scattering in a simple way and have made no attempt to account for water vapor absorption at 950~nm or any other atmospheric effects \citep{Shaw_2003}.

  \begin{figure}[htb]
      \includegraphics[width=80mm]{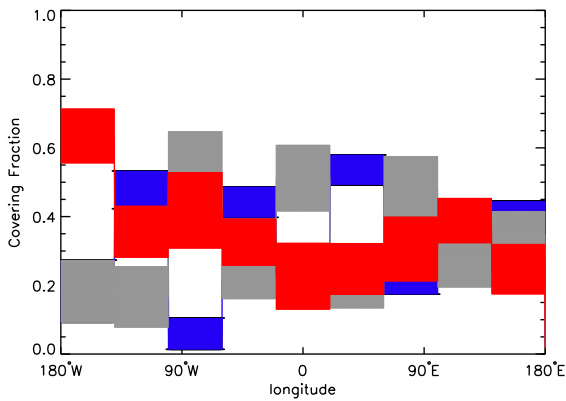}
    \caption{{\small Longitudinal maps of covering fraction, with vertical thickness of the lines corresponding to 1$\sigma$ uncertainty. The colors are the same as in Figure~\ref{pcp} and correspond to clouds (gray), ocean (blue), and land (red) viewed through a Rayleigh scattering atmosphere. \label{maps}}}
\end{figure}

The longitudinal maps of the three model surfaces (Figure~\ref{maps}) are not a good match to a cloud-free map of Earth. This is not surprising given our reliance on data from a single day, and the prevalence of obscuring clouds.  The large land content retrieved for the middle of the Pacific Ocean is spurious and may be due to our correction for changeable cloud cover (the observations begin and end with the spacecraft over the Pacific).  The right panel of Figure~\ref{pcp} hints at why the rotational map of Principal Component 1 was remarkably faithful to the actual surface geography of Earth \citep[Figure~10 of][]{Cowan_2009}: the land spectrum is pure PC1.\footnote{It should be noted, however, that \cite{Cowan_2009} performed PCA simultaneously on both \emph{Earth1} and \emph{Earth5} time series.} As noted in this paper, however, there is no reason to believe that ``maps'' of principal components should accurately reflect physical conditions on the planet.      

\section{Discussion}  
\begin{figure}[htb]
      \includegraphics[width=80mm]{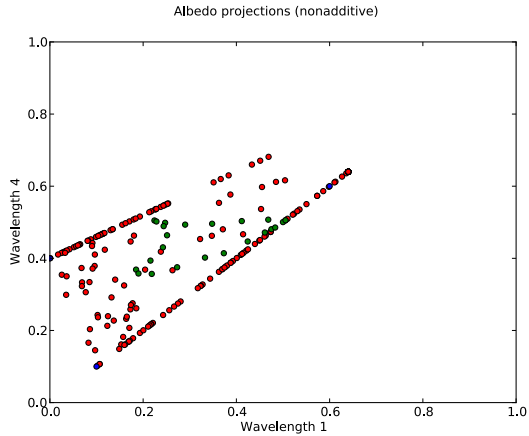}
    \caption{{\small Non-linear 3-surface toy model. The blue circles show the pure surface spectra: land $(0.0, 0.4)$, ocean $(0.1, 0.1)$, and cloud $(0.6,0.6)$.  The red circles show the colors of individual pixels, while green circles show the planet's apparent albedo path over a single planetary rotation. The covering fraction of land and ocean must sum to unity for every pixel, but for clouds we use a simple parameterization that captures the non-linear behavior of atmospheric scattering, $A = 1 - (1-A_{\rm cloud})(1-A_{\rm surf})$ , where $A_{\rm cloud}$  and $A_{\rm surf}$ are the cloud and surface albedos, respectively \citep{Cowan_2011}. Note that the pixel colors do not all lie within the triangle defined by the surface spectra, but are still amenable to endmember analysis. \label{nonlinear}}}
\end{figure}

The surface retrieval scheme presented here should be generally applicable, provided sufficient Euclidean distance between surfaces in color space (surfaces should look different), and large-amplitude variations in apparent covering fractions (large, longitudinally distinct geographical features). There are a number of well-justified assumptions made in the present work that should eventually be relaxed, however: \\
1) We assume Lambertian reflection for the purposes of convolving the planetary map of covering fractions with the visibility and illumination to obtain apparent covering fractions.  This assumption should be correct for the present data since they were obtained with Earth at slightly gibbous phase \citep{Williams_2008, Robinson_2010}.  In order to properly interpret data at crescent phases, however, it may be necessary to relax this assumption.\\ 
2) We adopt the known viewing geometry of Earth at the time of the observations. Any observations able to measure the rotational color-variations of an exoplanet would be more than adequate for estimating orbital phase and rotation period. The planetary obliquity and seasonal phase will not be known a priori, but simulations of full-orbit multiband lightcurves indicate that these geometrical parameters should be retrievable \citep{Kawahara_2010, Kawahara_2011, Fujii_2012}.\\   
3) We assume a static surface map for the planet, but in order to properly interpret weeks-months of data it would be imperative to allow the cloud cover to evolve in time. Clouds completely obscure the underlying surface in our linear model.  For example, 33\% cloud coverage at a given location means that that one-third of that pixel is covered in completely impenetrable clouds, while the remaining 67\% is perfectly cloud-free. Given this parametrization, changes in cloud cover necessarily involve changes in ocean and land coverage, and a significant increase in model complexity, all else being equal.\\
4) We assume that clouds combine linearly with actual surfaces. Optically thin clouds, however, obscure underlying surfaces while contributing to the reflectance spectrum, a non-linear effect \citep{Sabol_1992}. Numerical experiments indicate that the dimensionality of the apparent albedo locus is still $n_{\rm surf} -1$ for realistic, non-linear, radiative transfer \citep{Cowan_2011}.  Our experiments with a simple non-linear three-surface toy model further indicate that the apparent albedo locus is amenable to simplex shrink-wrapping, though none of the endmembers may correspond to a pure cloud spectrum (Figure~\ref{nonlinear}). In addition to being physically motivated, a non-linear cloud model would allow surface covering fractions to remain fixed despite changing cloud cover, enabling accurate surface maps with data spanning many planetary rotations. A major challenge with this approach would be determining, a priori, which surfaces should combine convexly and which should not.

5) Although in the present case the surface spectra were allowed to leave the principal component plane, it may be computationally necessary to constrain them to the PCP for larger $n_\lambda$ and/or $n_{\rm surf}$.\\  
6) We adopt three surfaces because the power spectrum of the albedo variations in dominated by the first two principal components. It is possible, however, that Earth has four or more surfaces that all happen to lie very close to the PCP. These surfaces might only betray themselves at higher spectral resolutions.  Putting aside that pathological case, it is plausible that some terrestrial exoplanets will have more than three major surface/cloud types, leading to a higher-dimensional locus in color space.  This would make the convex-hull and shrink-wrapping more computationally intensive, but existing algorithms are routinely applied to higher dimensional data \citep{Keshava_2003, Shaw_2003}. It is likely that the morphology of color variations in a higher-dimensional color space will still provide enough leverage to identify surface spectra, as was the case in the present study.

\section{Conclusions}
Our three-step surface-retrieval scheme consists of 1) performing principal component analysis on the multi-spectral reflectance matrix and projecting the data onto the principal component plane, 2) shrink-wrapping a simplex onto the projected data, and 3) relaxing the simplex vertices in order to match the time-variations in disk-integrated color.

Although we developed the method for directly-imaged terrestrial planets, there is no reason this method could not be used to determine the colors of clouds on directly-imaged gas giants, or the colors of albedo markings on unresolved Solar System bodies.

\acknowledgments 
W.M. Farr, B.W. Heumann, and N.A. Kaib provided valuable comments on early versions of this work. The EPOXI Mission of Opportunity is supported by the NASA Discovery Program.  The Deep Impact spacecraft was built by Ball Aerospace. The Earth observations, acquired as part of the EPOCh component of EPOXI, are available through MAST.

\end{document}